# Technology to scale up diversity in astronomy education

**Carmen Fies[1] and Chris Packham[1,2]**

[1]*The University of Texas at San Antonio*

*One UTSA Circle*

*San Antonio, TX 78249*

*USA*

[2] *National Astronomical Observatory of Japan,*

*2-21-1 Osawa,*

*Mitaka,*

*Tokyo, 181-8588,*

*Japan*

**Abstract**.  The San Antonio Teacher Training Astronomy Academy (SATTAA) completed its fourth annual iteration in June 2021 . While the program began as a face-to-face professional development opportunity for future and current school teachers, it transitioned to a fully online opportunity in 2020. In our efforts to offer an astronomy education program that is inclusive and particularly attentive to highly diverse populations, the transition to online programming became a core aspect of scaling up the program. The 2021 iteration featured an international facilitation team, and, for the first time, supported teachers from across the State of Texas. In this paper, we share data on how the facilitation team transitioned from a local to an international group, and on how the participant pool expanded from local to state-wide.

1.   **Background**

Astronomy can be perceived by some as esoteric and less relevant to our daily lives than other sciences. This is wholly incorrect, rather it is (1) highly inspirational, (2) arguably the most attractive STEM discipline, (3) closely aligned to astronautics and the space industry, and (4) tackles some of the most fundamental questions humankind has considered. As Carl Sagan once said, science is not a body of knowledge, it is a way of thinking to skeptically interrogate the world around us, including our leaders. At this turbulent time, a society equipped to make smart informed choices and avoid conspiracy theories is more important than ever.

This paper describes how we use astronomy's unique qualities as leverage to serve our local, regional and state-wide community, from K-12 education through to world-class research astrophysics, and especially underrepresented minority (URM) students. It also shows how, over time, the program grew and how the shift to fully online resulted in more diversity amongst facilitators and participants.

## 2.   Scaling up

We just completed our fourth year of SATTAA, a program we conceptualized as fully in person, and offered that way in 2018 and 2019. The outbreak of the pandemic in 2020 caused us to shift quickly and fully to online interactions. As reported elsewhere (i.e., Fies & Packham, 2021a, 2021b; Fies et al., 2021), even in the 2020 iteration, transitioning to fully online resulted in favorable outcomes in spite of rushed redesign. What stood out was the participants' high ranking of the experience on nearly all measures regardless of whether they participated in the f2f or online format. The 'forced' shift to fully online programming was a blessing in disguise for SATTAA as it helped to not only broaden the reach of the program to state-wide participation, but also to an international facilitation team.

### 2.1.   Facilitation Team

In all iterations, the facilitation team prepares to introduce each topic conceptually and to demonstrate hands-on methods to teach that topic. The level of content instruction is higher than needed for schools, so that the teachers are well equipped to answer questions from the students that go beyond what is required by the science standards of the state. All of the astrophysicists on the team were based in Texas in the 2018 and 2019 iterations; in 2020, with the transition to online learning, the team shifted to include one member from elsewhere in the country (see Table**). With more time to prepare, we were able to further diversify the team by including more national and now also international representation. In 2021, international experts from Spain, Chile, and Mexico facilitated sessions, exemplifying the international nature of astrophysics. This was noted by the participating teachers with excitement, which is particularly enriching as the facilitation team did not make any special announcements beyond the schedule itself.

Over the four years of SATTAA, the gender balance of the astrophysics facilitation team shifted. Where in 2018 more male than female members facilitated the pilot session, by 2019 equality was reached and then further shifted to more female than male representation (see Table1). This aligns with our goal to show diversity in the field. Although we did not explicitly draw attention to diversity, participants did remark on this in their feedback to us as well.

Table 1.   Astrophysics Facilitation Team; "local" signifies Texas

| YEAR | FEMALE | MALE | LOCAL | NATIONAL | INTERNATIONAL |
|---|---|---|---|---|---|
| **2018** | 3 | 5 | 9 | 0 | 0 |
| **2019** | 3 | 3 | 6 | 0 | 0 |
| **2020** | 5 | 2 | 6 | 1 | 0 |
| **2021** | 7 | 6 | 6 | 4 | 3 |

**2.2. Participants**

Since 2018, the program grew from a local teacher professional development program to a program with state-wide participation. From a pilot year that included only secondary preservice teachers (students at UTSA who studied to become STEM teachers), the program grew to include current secondary school teachers, future elementary school teachers, and current elementary school teachers. Teachers from school districts spanning Texas east-to-west and north-to-south joined SATTAA2021; they teach at urban and rural schools, and in mostly less affluent districts.

The majority of participants (48%) are current STEM teachers in secondary schools, followed by future secondary STEM teachers (29%). Similarly, of the primary school teachers, more were already teaching (18%) than preparing to teach (5%) in elementary school settings (Figure 1).

Figure 1: Percent distribution of participants. Note that the percent distribution in the chart below is based on percent of total participants in all four years of the program.

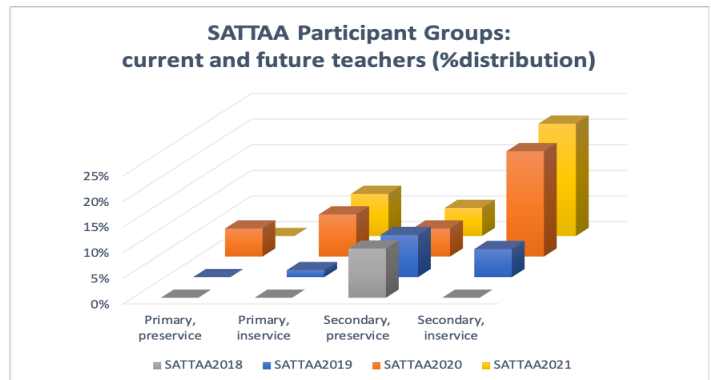

An important aspect of SATTAA is to support teachers who work predominantly with minority and economically disadvantaged learners. The data show that, in the three years of in-service teacher participation, the participants' school districts were home to student populations that are overwhelmingly non-white and substantively economically disadvantaged. Focusing on San Antonio school districts, our point of comparison is that of the city-wide population (United States Census Bureau, 2021) and summarized in Table 2 and Figure 2. In comparison, SATTAA2021 teachers who work at schools elsewhere in the State of Texas teach students who are more likely to be white (40.7%) and somewhat more affluent (36% economically disadvantaged).

Table 2 and Figure 2. Participating San Antonio teachers' school district demographics.

|  | *San Antonio* | *SATTAA2019* | *SATTAA2020* | *SATTAA2021* |
|---|---|---|---|---|
| *% White* | 24.7 | 15.3 | 11.4 | 16.5 |
| *% Economically Disadvantaged* | 17.8 | 55.9 | 62.9 | 67.1 |

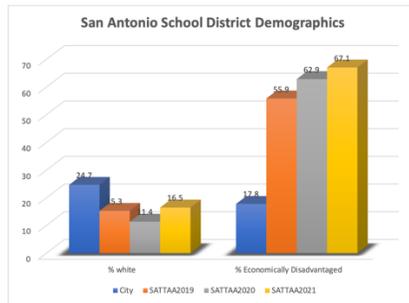

## 3. The Role of Technology

Although the pandemic caused great disruption to every aspect of life, on the balance, shifting to fully online teacher professional development resulted in benefits to SATTAA. While we lost some of the immediacy of working in the same space and the opportunity to visit exciting real locations, we gained equality in opportunity, albeit with a caveat. We are aware that the equality in opportunity we gained is flawed. School funding is inequitable and technologies available to teachers are therefore also disparate. Moreover, many teachers have to work in other jobs during the summers to make ends meet. The districts our participants work in tend to span the gap between more and less affluent, albeit more heavily leaning toward the less affluent. We do not know how many teachers might have wished to participate, but could not because of technology requirements or because they could not afford the time away from work. To help mediate this gap, we are working on identifying funds to pay participating teachers stipends for participation.

As described above, SATTAA became more diverse in race, gender, geography and language. Although that diversification is a programmatic goal, achieving this so quickly was directly aided by moving to online summers. We were able to attract co-facilitators who otherwise could not have participated because of added time for travel, and who we might not have had funds to arrange the travel for. We also now support teachers from urban and rural school districts state-wide, many of whom would not be able to join an in-person program in San Antonio.

Logistics in general have a smaller impact in the online years, resulting in teachers working on honing their knowledge and skills for a larger part of the total time (Fies & Packham, 2021b). This largely also is related to the fact that travel is not necessary. Overall, the program has much lower budgetary needs since the 2020 shift to online sessions. Whereas the first two years required substantive funding for materials and field trips, the program cost now is limited to a single fee associated with one of the virtual field trips. Since the program was, and continues to be, offered free-of-charge to participants in every year, identifying funds is a much less critical component now in general.

We also use digital technologies to support the SATTAA Alumni group, a community of practice (Fies & Packham, 2021a). Alumni receive weekly updates, send questions and share events. Alumni also are eligible to receive grants for astronomy education in their own classrooms. These grants are funded through a local donor.

SATTAA online has a much-reduced carbon footprint (Fies & Packham, 2021a). The extent of reduced emissions can be estimated and offset against increased network traffic and server activity. The results of this analysis, and at estimate of the context, will be published in the future.

## 4.  Impact Evaluation

The feedback we receive from participants is important to us. However, we evaluate the outcomes of SATTAA each year not only in connection with the participants' experiences in our sessions (Fies & Packham, 2021a, 2021b; Fies et al., 2021), but also collect data to evaluate the impact on students in K-12 classrooms.

In order to estimate the 5-year impact the program has on K-12 learners, we assume that, from 2022-2026, each teacher teaches five classes in an academic year, that 25 students are in each of these classes, and that the 2018-2020 preservice teachers are in classrooms. Even if SATTAA never took place again, roughly 40,000 students in the San Antonio Metro area, and 6,875 students elsewhere in Texas would benefit from their teachers' participation. However, we expect the program to continue with an estimated participation of 32 teachers in each year, which will add another 60,000 students to benefit by 2026.

## 5.  Conclusion

When we first began to design SATTAA, we agreed that the program should be free-of-charge to participating teachers, that it should be facilitated by an interdisciplinary team of experts and connect astronomy content, pedagogy, and technology explicitly. Our goals are to positively contribute to improved and inclusive STEM education, with associated improvement in STEM interest and STEM literacy, and to focus in particular on teachers at schools in need. Our data show that the shift to fully online programming has moved us closer to achieving these goals at a faster pace than anticipated. That said, we see gaps that need to be addressed swiftly and decisively, such as finding funding for teacher stipends, to further improve equitable access.

**Acknowledgements**. We thank NASA, the Space Telescope Science Institute, and NSF grant no. 1616828 for supporting this work financially. We also gratefully acknowledge the generous support of BEAT LLC (https://beatllc.com) for providing the assistance to enable SATTAA grants starting 2020.